\documentclass[aps,pra,prabib,twocolumn,showpacs,nofootinbib]{revtex4}
\usepackage{graphicx} \usepackage{amsmath} \usepackage{amssymb}
\usepackage{amsfonts} \usepackage{bm}

\begin{document}

\newcommand{\be}{\begin{equation}} \newcommand{\ee}{\end{equation}}
\newcommand{\bea}{\begin{eqnarray}}\newcommand{\eea}{\end{eqnarray}}

\title{Non-hermitian models in higher dimensions}

\author{Pulak Ranjan Giri} \email{pulakranjan.giri@saha.ac.in}

\affiliation{Theory Division, Saha Institute of Nuclear Physics,
1/AF Bidhannagar, Calcutta 700064, India}

\begin{abstract}
It is know that $PT$-symmetric models have real spectra provided the
symmetry is not spontaneously broken. Even pseudo-hermitian models
have real spectra, which enlarge the the class of non-hermitian
models possessing  real spectra.  We however  consider non-hermitian
models in higher dimensions which are not necessarily explicit
$PT$-symmetric nor pseudo-hermitian. We show that the models may
generate real spectra depending upon the coupling constants of the
Hamiltonian. Our models thus further generalize the class of
non-hermitian systems, which possess real spectra.
\end{abstract}


\date{\today}

\maketitle

The spectrum of a quantum mechanical system  becomes real if the
Hamiltonian, $H$,  of the corresponding system is self-adjoint
\cite{reed}. In terms of mathematics, for a self-adjoint
Hamiltonian, the domain for it, $D(H)$, is equal to the domain,
$D(H^\dagger)$, of its corresponding adjoint operator, $H^\dagger$.
This constraint implies that the Hamiltonian would have to be
hermitian, $\int(\psi_1)^\dagger H\psi_2=
\int(H\psi_1)^\dagger\psi_2,~\forall \psi_1,\psi_2\in D(H)$.
However,  in order to have real spectra it is not necessary for the
Hamiltonian to be hermitian. For example, it is known that for a
non-hermitian system we may have real spectra if the Hamiltonian is
$PT$-symmetric \cite{bender1} and if the symmetry is not
spontaneously broken. Complex eigenvalues may arise if the
$PT$-symmetry is spontaneously broken. Even the the constraint,
$PT$-symmetry, can be relaxed to include larger class of models
known as pseudo-hermitian, defined by the relation $H^\dagger=
SHS^{-1}$, for an hermitian operator $S$. For pseudo-hermitian
Hamiltonian there exists a corresponding hermitian Hamiltonian which
has identical spectrum.

The discussion on non-hermitian quantum mechanics is hugely
discussed in  one dimension
\cite{bender2,bender3,bender4,sinha,dorey1,dorey2,bender5,ali1,ali2,ali3}. 
One can however
generalize the non-hermitian models to higher dimensions. For
example, the higher dimensional $PT$-symmetric potentials
\cite{levai,asiri,habib} has been studied, which renders real
spectra. In spherical coordinates $PT$-symmetry is defined by
$PTr{PT}^{-1}= r, PT\theta {PT}^{-1}= \pi-\theta, PT\phi {PT}^{-1}=
\pi +\phi$, $PTi{PT}^{-1}= -i$. For a Hamiltonian, $H$, which is
separable in spherical coordinates one can write it as $H= H_r +
H_\Omega$, where $H_r$ is the radial part and $H_\Omega$ is the
remaining angular part of the Hamiltonian. Then the $PT$-symmetry of
the Hamiltonian can be decomposed as $PTH{PT}^{-1}= PTH_r{PT}^{-1} +
PTH_\Omega{PT}^{-1}$. Since the radial coordinate $r$ remains
invariant under $PT$-transformation, potential of $H_r$ should be
real if it respects the symmetry, $PTH_r{PT}^{-1}= H_r$. However
$PTH_\Omega{PT}^{-1}= H_\Omega$ implies that the potential
$V(\theta,\phi)$  for  $H_\Omega$ may be complex, but would have to
be invariant under $PT$-transformation, $PTV(\theta,\phi){PT}^{-1}=
V(\theta,\phi)$.

One can even think about the relaxation of the $PT$-symmetry to
include more potentials in higher dimensions, which may have real
spectra. But the question is whether it is necessary for a higher
dimensional systems to be at least pseudo-hermitian  in order to
render real eigenvalues. We in this article focus on this issue by
studying some general complex potentials. The answer will not be
given but the problem we study will give a first step forward in
this direction. One should note that it has been shown in
\cite{kumar} that the Hamiltonian for the Calogero model with
complex coupling of the inverse square potential does possess real
eigen-value if the coupling is chosen properly. Although the
Calogero model is in one-dimension, it can be thought of as a higher
dimensional model if the co-ordinates of the $N$ Calogero particles
on a line are thought to be co-ordinates of a single particle in
higher dimensions. Calogero model with complex coupling is a
specific example, which possesses real spectrum. But in a general
framework it is also possible to show that a complex potential may
possesses real spectra. It can also be recalled that in
\cite{pulak2} a scale invariant non-hermitian but $PT$-symmetric
potential $V_{\mathcal{P}\mathcal{T}}=\frac{c}{r(r+z)}+
\frac{c^*}{r(r-z)}$  was considered, in which an electrically
charged particle is moving in the background field of a magnetic
monopole. This system can be obtained from the generalized
MIC-Kepler model, which is the system of two dyons with the axially
symmetric potential $V_{MIC}= \frac{c_1}{r(r+z)}+
\frac{c_2}{r(r-z)}- \alpha_s/r + s^2/r^2$,  by setting  the Coulomb
term and the extra inverse square term zero, i.e., $\alpha_s= s^2=0$
and and generalizing  the two constants  $c_1$ and $c_2$ to complex
conjugate numbers. It was  shown that the electrically charged
particle possesses bound states.

In this article, we discuss in a  generic framework that   a model
in $N$-dimensions ($N\geq 2$) interacting with the complex
anisotropic inverse square potential
\begin{eqnarray}
V(r,\theta,\phi_i)= \frac{1}{r^2}\left(\alpha+
\beta\mathcal{F}(\theta,\phi_i)\right)\,, \label{pot}
\end{eqnarray}
where $\alpha$ and $\beta$ are complex couplings,  has real
spectrum. The advantage of this potential is that the corresponding
Hamiltonian can be separated in radial and angular part. We keep the
function $\mathcal{F}(\theta,\phi_i)$  completely general and do not
assume any specific form at this stage. Although this potential is
general,  it is not completely arbitrary. Due to the form of the
potential it makes the Hamiltonian scale invariant.   Note however
that our potential (\ref{pot}) is complex and may not possess
$PT$-symmetry in general. Then the one particle Hamiltonian  reads
\begin{eqnarray}
H=-\frac{\hbar^2}{2\mu}\nabla + \frac{1}{r^2}\left(\alpha+
\beta\mathcal{F}(\theta,\phi_i)\right)\,. \label{ham}
\end{eqnarray}
Here $\mu$ is the reduced mass of the particle. Note that although
(\ref{ham}) is written in a general form, it can be related to the
electron polar molecule system with the dipolar molecules are taken
to be point like. For $\alpha=0$, $\beta=eD$(real) and
$\mathcal{F}(\theta,\phi_i)=\cos\theta$, (\ref{ham}) reduces to the
Hamiltonian of electron polar molecule system \cite{pulak1}. In
spherical polar coordinates  (\ref{ham}) can be separated in radial
and angular co-ordinates as
\begin{eqnarray}
H= H_r \oplus r^{-2}H_\Omega \,.\label{hamoplus}
\end{eqnarray}
Due to the existence of the complex potential (\ref{pot}), now both
the Hamiltonians
\begin{eqnarray}
H_r &=& -\frac{h^2}{2\mu}\frac{d^2}{dr^2}-
\frac{h^2}{2\mu}\frac{N-1}{r}\frac{d}{dr}
+\frac{\alpha}{r^2}\label{ham1}\\ H_\Omega &=&
-\frac{h^2}{2\mu}\nabla_\Omega +\beta\mathcal{F}(\theta,\phi_i)
\label{ham2}
\end{eqnarray}
are non-hermitian and possess no $PT$-symmetry. Here the Hilbert
space in $N$-dimensions, $L^2(\mathbb{R}^N, r^{N-1}dr d\Omega)$, has
been separated in two orthogonal spaces as  $L^2(\mathbb{R}^+,
r^{N-1}dr)\otimes L^2(S^{N-1},d\Omega)$. The eigenvalue equation
$H\Psi= E\Psi$ now can be written as two eigenvalue equations in
mutually orthogonal Hilbert spaces
\begin{eqnarray}
\label{ham31}H_{r,\eta}R_{E,\eta}(r) &=&ER_{E,\eta}(r),\\H_\Omega
Y_\eta(\theta,\phi_i)&=&\eta Y_\eta(\theta,\phi_i)\,, \label{ham3}
\end{eqnarray}
where
\begin{eqnarray}
H_{r,\eta}= -\frac{h^2}{2\mu}\frac{d^2}{dr^2}-
\frac{h^2}{2\mu}\frac{N-1}{r}\frac{d}{dr}+
\frac{(\alpha+\eta)}{r^{2}}\,. \label{ham4}
\end{eqnarray}
Since the eigenvalue equation for $H_\Omega$ is non-hermitian and
does not possess any constraint like $PT$-symmetry or
pseudo-hermiticity, its eigenvalue $\eta$ will be complex in general
and depends on the complex coupling constant $\beta$. Suppose $\eta=
\eta_R + i\eta_I$ and $\alpha=\alpha_R + i\alpha_I$. We now  choose
a  specific form of the coupling constant $\alpha$ such that
$\alpha_I= -\eta_I$. Then the two complex terms  of the inverse
square potential cancels each other and only the real terms survive,
$(\eta_R + \alpha_R)r^{-2}$. The effective radial Hamiltonian
\begin{eqnarray}
H_{r,\eta_R}= -\frac{h^2}{2\mu}\frac{d^2}{dr^2}-
\frac{h^2}{2\mu}\frac{N-1}{r}\frac{d}{dr}+
\frac{(\alpha_R+\eta_R)}{r^{2}}\,, \label{ham5}
\end{eqnarray}
is now hermitian and does possess real eigenvalue but whether it has
bound states or not depends on the ranges of the coupling constant
$\alpha_R+\eta_R$. The solution is however well known \cite{pulak3}.
We give a brief discussion  here for completeness. One can now use
the transformation $R_{E,\eta}(r)= r^{-(N-1)/2}\chi_{E,\eta}(r)$ on
the Schr\"{o}dinger eigenvalue equation $H_{r,\eta_R}R_{E,\eta}(r)=
ER_{E,\eta}(r)$. The Hamiltonian of the transformed eigenvalue
equation $\mathcal{H}_{r,\eta_R}\chi_{E,\eta}(r)= E\chi_{E,\eta}(r)$
has the very familiar form $\mathcal{H}_{r,\eta_R}= -\partial^2_r+
\mbox{g}/r^2$, with $\mbox{g}= l(l+N-2) + \alpha_R + \eta_R+
(N-1)(N-3)/4$. It can be shown that $\mathcal{H}_{r,\eta_R}$ have
only one bound state for $-1/4\leq\mbox{g}<3/4$. The bound state
solutions can be found from the self-adjoint  domain
\begin{eqnarray}
\mathcal{D}^\Sigma(L^{-2})=  \{\mathcal{D} +
\psi(r,L^{-2},\Sigma)|\psi(r,L^{-2},\Sigma)\in \mathcal{D}\}\,,
\label{NHds}
\end{eqnarray}
where
$\mathcal{D}=\{\psi(r)\in\mathcal{L}^2(r^{N-1}
dr),\psi(0)=\psi'(0)=0\}$
and $\mathcal{D}$ is the domain of the adjoint Hamiltonian
${\mathcal{H}_{r,\eta_R}}^*$. The length  scale $L$ in the domain
$\mathcal{D}^\Sigma(L^{-2})$ comes from the self-adjoint extension
mechanism. The energy eigenvalue is
\begin{eqnarray}
\nonumber  E &=& -L^{-2}\mathcal{F}_1(\Sigma),~~ g\neq-1/4~,\\
&=&-L^{-2}\mathcal{F}_2(\Sigma),~~ g= -1/4\,. \label{2b}
\end{eqnarray}
The exact form of the functions $\mathcal{F}_{1,2}(\Sigma)$ can be
found from the matching condition of the eigen-function with the
domain $\mathcal{D}^\Sigma(L^{-2})$. The bound state eigenfunction
is of the form \cite{pulak3}
\begin{eqnarray}
\nonumber R_{E,\eta}(r) &\equiv & K_{\sqrt{g+1/4}}(\sqrt{E}r),~~ g\neq-1/4~,\\
&\equiv & K_0(\sqrt{E}r),~~ g=-1/4~\,,\label{2bs}
\end{eqnarray}
where $K$s are  modified Bessel function.  For more attractive
coupling constant $g <-1/4$, the system  does possess infinitely
many bound states and goes up to negative infinity. For more
positive coupling constant $g\geq 3/4$ the system is essentially
self-adjoint and does not possess bound state. We now move to some
specific examples to justify the above discussion.

{\it 2-dimensional model:} We consider a 2-dimensional system with
complex non $PT$-symmetric potential
\begin{eqnarray}
V(r,\phi)= \frac{1}{r^2}\beta \exp(i\phi)\,. \label{pot1}
\end{eqnarray}
Note that this potential belongs to the class defined in (\ref{pot})
where now $\alpha=0$ and $\mathcal{F}(\theta,\phi_i)=\exp(i\phi)$
have been chosen.  For real $\beta$ this potential coincides with
the potential used in \cite{habib}.  The angular  eigenvalue
equation (\ref{ham3}) can be solved exactly and it has two
independent solutions in terms of modified bessel functions (we set
the units $\hbar= 2\mu=1$ through out)
\begin{eqnarray}
\label{azisol1}Y_\eta(\phi)\equiv
I_{2\sqrt{\eta}}(2\sqrt{\beta}\exp(i\phi/2))\,,\\
\equiv
K_{2\sqrt{\eta}}(2\sqrt{\beta}\exp(i\phi/2))\,.\label{azisol2}
\end{eqnarray}
To find out the exact solution and the eigenvalue $\eta$ we use the
standard periodicity boundary condition of the wave-function for the
azimuthal variable $Y_\eta(\phi)= Y_\eta(\phi+2\pi)$, which rejects
the second solution $K$. The solution thus becomes (\ref{azisol1})
with $\sqrt{\eta}$  being non negative integers. Note that although
the potential (\ref{pot1}) is non $PT$-symmetric and complex it
renders real eigenvalues.  The radial part of the eigenvalue
equation obtained from (\ref{ham31}),
\begin{eqnarray}
\left[-\frac{d^2}{dr^2}- \frac{1}{r}\frac{d}{dr}+
\frac{\eta}{r^{2}}\right]R_{E,\eta}(r)= ER_{E,\eta}(r)\,,
\label{2dradial}
\end{eqnarray}
is hermitian and it has real eigenvalues once the boundary condition
is chosen properly. The solutions can be obtained from (\ref{2b})
and (\ref{2bs}). This example is thus a generalization of the class
of non-hermitian quantum mechanical systems.

{\it 3-dimensional model:} For the 3-dimensional case we consider a
potential of the form
\begin{eqnarray}
V(r,\theta,\phi)= \frac{1}{r^2}\left(\alpha+ \beta\cot^2\theta
+\gamma\exp(i\phi)\right)\,. \label{pot3}
\end{eqnarray}
Note that  this type of potential with $\gamma=0$ can be found in
\cite{dong}, but the difference is we have made all the coupling
constants complex.   The eigenvalue problem for the angular variable
is now
\begin{eqnarray}
\left[-\frac{1}{\sin\theta}\left(\sin\theta\frac{d}{d\theta}\right)
+ \frac{\beta\cos^2\theta + m^2}{\sin^2\theta}\right]Y_{\eta}= \eta
Y_{\eta}\,. \label{angleham1}
\end{eqnarray}
This equation is exactly solvable. For real $\beta$ and $\gamma=0$,
it has been solved in \cite{dong} with eigenvalue $\eta$ being real.
For complex coupling constants  the solutions can be written as
\begin{eqnarray}
Y_\eta(\theta,\phi)\equiv
P_\mu^\nu(\cos\theta)I_{2m}(2\sqrt{\gamma}\exp(i\phi/2))\,,
\label{legendre1}
\end{eqnarray}
where $\nu= \sqrt{\beta+m^2}$ and $\mu(\mu+1)=\eta +\beta$, but the
eigenvalue  $\eta$ is now complex in general. $P^\nu_\mu$ is the
legerdre  function and $I_{2m}$ is the modified bessel function.
This time the radial Hamiltonian is non-hermitian due to the complex
potential $(\alpha +\eta)r^{-2}$. But as mentioned before we can
choose the complex part of the coupling $\alpha$ such that $\alpha_I=
-\eta_I$. This will make the potential $(\alpha +\eta)r^{-2}=
(\alpha_R +\eta_R)r^{-2}$ real. The effective radial eigenvalue
equation,
\begin{eqnarray}
\left[-\frac{d^2}{dr^2}- \frac{2}{r}\frac{d}{dr}+ \frac{\alpha_R+
\eta_R}{r^{2}}\right]R_{E,\eta}(r)= ER_{E,\eta}(r)\,,
\label{3dradial}
\end{eqnarray}
thus becomes hermitian and its solution can be  obtained from
(\ref{2b}) and (\ref{2bs}). Note that it is also an example of
non-hermitian model which generalizes the class of non-hermitian
quantum mechanical systems admitting real eigenvalues.

In conclusion, we discussed non-hermitian models in higher
diemsnions. The problems discussed are not necessarily
$PT$-symmetric nor explicitly  pseudo-hermitian.  Exploiting the
complex eigen-values of the angular Hamiltonian we managed to cancel
the complex potential of the radial Hamiltonian, which makes it
hermitian. Although in order to keep the eigenvalue real a
Hamiltonian must be pseudo-hermitian at least, in our discussion the
higher dimensional Hamiltonians are not explicitly  so, still giving
real spectra. More efforts are needed to answer the question, raised
above, that whether a system should have to be at least
pseudo-hermitian in order to get the spectra real. However it may be
possible that for a Hamiltonian $H$, which is non $PT$-symmetric and
non pseudo-hermitian and thus having complex eigen-value $E_R + iE_I$,
one can construct a  Hamiltonian $H-iE_I$, which will become
$PT$-symmetric or pseudo-hermitian and therefore will generate real spectra.

\end{document}